\documentclass[sigconf]{aamas}  

\usepackage{booktabs}
\usepackage{amsmath,amssymb,bm}
\usepackage{amsfonts}
\usepackage{amsthm}
\usepackage{sgame, tikz}
\usepackage{subcaption}
\usepackage{graphicx}
\usepackage{hyperref}

\setcopyright{ifaamas}  
\acmDOI{doi}  
\acmISBN{}  
\acmConference[AAMAS'18]{Proc.\@ of the 17th International Conference on Autonomous Agents and Multiagent Systems (AAMAS 2018), M.~Dastani, G.~Sukthankar, E.~Andre, S.~Koenig (eds.)}{July 2018}{Stockholm, Sweden}  
\acmYear{2018}  
\copyrightyear{2018}  
\acmPrice{}  

\newcommand\secspace{0.3}

\begin{document}

\title{A Generalised Method for Empirical Game Theoretic Analysis}  




%
\author{Karl Tuyls}
\affiliation{%
  \institution{DeepMind}
  \streetaddress{6 Pancras Square}
  \city{London, UK} 
}
\email{karltuyls@google.com}
\author{Julien Perolat}
\affiliation{%
  \institution{DeepMind}
  \streetaddress{6 Pancras Square}
  \city{London, UK} 
}
\email{perolat@google.com}

\author{Marc Lanctot}
\affiliation{%
  \institution{DeepMind}
  \streetaddress{6 Pancras Square}
  \city{Edmonton, Canada} 
}
\email{lanctot@google.com}

\author{Joel Z Leibo}
\affiliation{%
  \institution{DeepMind}
  \streetaddress{6 Pancras Square}
  \city{London, UK} 
}
\email{jzl@google.com}

\author{Thore Graepel}
\affiliation{%
  \institution{DeepMind}
  \streetaddress{6 Pancras Square}
  \city{London, UK} 
}
\email{thore@google.com}

\begin{abstract}  
This paper provides theoretical bounds for empirical game theoretical analysis of complex multi-agent interactions. We provide insights in the empirical meta game showing that a Nash equilibrium of the meta-game is an approximate Nash equilibrium of the true underlying game. We investigate and show how many data samples are required to obtain a close enough approximation of the underlying game. Additionally, we extend the meta-game analysis methodology to asymmetric games. The state-of-the-art has only considered empirical games in which agents have access to the same strategy sets and the payoff structure is symmetric, implying that agents are interchangeable. Finally, we carry out an empirical illustration of the generalised method in several domains, illustrating the theory and evolutionary dynamics of several versions of the \textit{AlphaGo} algorithm (symmetric), the dynamics of the Colonel Blotto game played by human players on Facebook (symmetric), and an example of a meta-game in Leduc Poker (asymmetric), generated by the PSRO multi-agent learning algorithm.
\end{abstract}

%


\maketitle

\vspace{-0.5cm}
\section{Introduction}

\noindent Using game theory to examine multi-agent interactions in complex systems is a non-trivial task. Works by Walsh et al. \cite{Walsh02,Walsh03} and Wellman et al. \cite{Wellman06}, have shown the great potential of using heuristic strategies and empirical game theory to examine such interactions at a higher meta-level, instead of trying to capture the decision-making processes at the level of the atomic actions involved. Doing this turns the interaction in a smaller normal form game, or meta-game, with the higher-level strategies now being the primitive actions of the game, making the complex multi-agent interaction amenable to game theoretic analysis.

\noindent Others have built on this empirical game theoretic methodology and applied these ideas to no limit Texas hold'em Poker and various types of double auctions for example, see \cite{PhelpsPM04,PonsenTKR09,PhelpsCMNPS07,KaisersTTP08,TuylsP07}, showing that a game theoretic analysis at the level of meta-strategies yields novel insights into the type and form of interactions in complex systems.

\noindent A major limitation of this empirical game theoretic approach is that it comes without theoretical guarantees on the approximation of the true underlying game by an estimated game based on sampled data, and that it is unclear how many data samples are required to achieve a good approximation. Additionally, the method remains limited to symmetric situations, in which the agents or players have access to the same set of strategies, and are interchangeable. One approach is to ignore asymmetry (types of players), and average over many samples of types resulting in a single expected payoff to each player in each entry of the meta-game payoff table. Many real-world situations though are asymmetric in nature and involve various roles for the agents that participate in the interactions. For instance, buyers and sellers in auctions, or games such as Scotland Yard, but also different roles in e.g. robotic soccer (defender vs striker) and even natural language (hearer vs speaker). 

\noindent In this paper we tackle these problems. We prove that a Nash equilibrium of the estimated game is a $2 \epsilon$-Nash equilibrium of the real underlying game, showing that we can closely approximate the real Nash equilibrium as long as we have enough data samples from which to build the meta-game payoff table. Furthermore, we also examine how much data samples are required to confidently approximate the underlying game.  We also show how to generalise the heuristic payoff or meta-game method introduced by Walsh \textit{et al.} to two-population asymmetric games.

\noindent Finally, we illustrate the generalised method in several domains. We carry out an experimental illustration on the \textit{AlphaGo} algorithm \cite{DSilverHMGSDSAPL16}, Colonel Blotto \cite{KohliKBHSG12} and an asymmetric Leduc poker game. In the \textit{AlphaGo} experiments we show how a symmetric meta-game analysis can provide insights into the evolutionary dynamics and strengths of various versions of the \textit{AlphaGo} algorithm while it was being developed, and how intransitive behaviour can occur by introducing a non-related strategy. In the Colonel Blotto game we illustrate how the methodology can provide insights into how humans play this game, constructing several symmetric meta-games from data collected on Facebook. Finally, we illustrate the method in Leduc poker, by examining an asymmetric meta-game, generated by a recently introduced multiagent reinforcement learning algorithm, policy-space response oracles (PSRO) \cite{Lanctot17}. For this analysis we rely on some theoretical results that connect an asymmetric normal form game to its symmetric counterparts \cite{TuylsSym}.

 

\vspace{-\secspace cm}
\section{Preliminaries}\label{sec:prelim}

In this section, we introduce the necessary background to describe our game theoretic meta-game analysis of the repeated interaction between $p$ players. 

\vspace{-\secspace cm}
\subsection{Normal Form Games:} In a $p$-player Normal Form Game (NFG), players are involved in a single round strategic interaction. Each player $i$ chooses a strategy $\pi^i$ from a set of $k$ strategy $S^i = \{\pi_1^i, \dots,\pi_k^i\}$ and receives a payoff $r^i(\pi^1, \dots, \pi^p): S^1 \times \dots \times S^p \rightarrow \mathbb{R}$. For the sake of simplicity, we will write $\bm{\pi}$ the joint strategy $(\pi^1,\dots,\pi^p)$ and $\bm{r}(\bm{\pi})$ the joint reward $(r^1(\bm{\pi}), \dots,r^p(\bm{\pi}))$. Then a $p$-player NFG is a tuple $G=(S^1, \dots, S^p, r^1, \dots, r^p)$. 
Each player interacts in this game by following a strategy profile $x^i$ which is a probability distribution over $S^i$.

A symmetric NFG captures interactions where players can be interchanged. The first condition is therefore that the strategy sets are the same for all players, (\textit{i.e.} $\forall i,j \; S_i=S_j$ and will be written $S$). In a symmetric NFG, if a permutation is applied to the joint strategy $\bm{\pi}$, the joint payoff is permuted accordingly. Formally, a game $G$ is symmetric if for all permutations of $p$ elements $\sigma$ we have $\bm{r}(\bm{\pi}_{\sigma}) = \bm{r}_{\sigma}(\bm{\pi}) $ (where $\bm{\pi}_{\sigma} = (\pi^{\sigma(1)}, \dots, \pi^{\sigma(p)})$ and $\bm{r}_{\sigma}(\bm{\pi}) = (r^{\sigma(1)}(\bm{\pi}), \dots,r^{\sigma(p)}(\bm{\pi}))$). So for a game to be symmetric there are two conditions, the players need to have access to the same strategy set and the payoff structure needs to be symmetric, such that players are interchangeable. If one of these two conditions is violated the game is asymmetric.

In the asymmetric case our analysis will focus on the two-player case (two roles) and thus we introduce specific notations for the sake of simplicity. In a two-player normal-form game, each player's payoff can be seen as a $k \times k$ matrix. We will write $A = (a_{l,l'})_{1 \leq l,l' \leq k}$ for the payoff matrix of player one (\textit{i.e.} $a_{l,l'} = r^1(\pi^1_l, \pi^2_{l'})$) and $B = (b_{l,l'})_{1 \leq l,l' \leq k}$ for the payoff matrix of player two  (\textit{i.e.} $b_{l,l'} = r^2(\pi^1_l, \pi^2_{l'})$). In this two-player game, the column vector $x$ is the strategy of player one and $y$ the one of player two. In the end, a two player NFG is defined by the following tuple $G=(S^1, S^2, A, B)$.

\vspace{-\secspace cm}
\subsection{Nash Equilibrium}
In a two-player game, a pair of strategies $(x,y)$ is a Nash equilibrium of the game $(A,B)$ if no player has an incentive to switch from their current strategy. In other words, $(x,y)$ is a Nash equilibrium if $x^\top A y = \max Ay$ and $x^\top B y = \max x^\top B$.

Evolutionary game theory often consider a single strategy $x$ that plays against itself. In this situation, the game is said to have a single population. In a single population game, $x$ is a Nash equilibrium if $x^\top A x = \max Ax$.
\vspace{-\secspace cm}
\subsection{Replicator Dynamics}
\label{ReplicatorDynamics}

The replicator dynamics equation describes how a strategy profile evolves in the midst of others. This evolution is described according to a first order dynamical system. In a two-player NFG $(A, B, S^1, S^2)$, the replicator equations are defined as:
\vspace{-0.2cm}
\begin{align}\label{eq:asymRD1}
    &\dot{x}_l = x_l \left( (A y)_l - x^\top A y\right) 
    &\dot{y}_{l'} = y_{l'} \left( (x^\top B)_{l'} - x^\top B y \right) 
\end{align}

The dynamics defined by these two coupled differential equations changes the strategy profile to increase the probability of the strategies that have the best return or are the \textit{fittest}.

In the case of a symmetric two-player game ($A=B^\top$), the replicator equations assume that both players play the same strategy profile (\textit{i.e.} player one and two play according to $x$) and the dynamics is defined as follows:
\vspace{-0.2cm}
\begin{align}\label{eq:singleRD}
    &\dot{x}_l = x_l \left( (A x)_l - x^\top A x\right)
\end{align}

\vspace{-\secspace cm}
\subsection{Meta Games}

A meta game is a simplified model of a complex interaction. In order to analyze complex games like e.g. poker, we do not need to consider all possible strategies but a set of relevant meta-strategies that are often played~\cite{PonsenTKR09}. These meta strategies (or styles of play), over atomic actions, are commonly played by players such as for instance "passive/aggressive" or "tight/loose" in poker. A $p$-type meta game is now a $p$-player repeated NFG where players play a limited number of meta strategies. Following our poker example, the strategy set of the meta game will now be defined as the set $S=\{\textbf{"aggressive"}, \textbf{"tight"}, \textbf{"passive"}\}$ and the reward function as the outcome of a game between $p$-players using different profiles.

\vspace{-\secspace cm}
\section{Method}\label{sec:method}


\noindent There are now two possibilities, either the meta-game is symmetric, or it is asymmetric. We will start with the simpler symmetric case, which has been studied in empirical game theory, then we continue with asymmetric games, in which we consider two populations, or roles.

\vspace{-\secspace cm}
\subsection{Symmetric Meta Games}
We consider a set of agents or players $A$ with $|A|=n$ that can choose a strategy from a set $S$ with $|S|=k$ and can participate in one or more $p$-type meta-games with $p \leq n$. 
If the game is symmetric then the formulation of meta strategies has the advantage that the payoff for a strategy does not depend on which player has chosen that strategy and consequently the payoff for that strategy only depends on the composition of strategies it is facing in the game and not on who is playing the strategy. This symmetry has been the main focus of the use of empirical game theory analysis \cite{Walsh02,Wellman06,PonsenTKR09,PhelpsCMNPS07}.



\noindent If we were to construct a classical payoff table for $\mathbf{r}$ we would require $k^p$ entries in the table (which becomes large very quickly). Since all players can choose from the same strategy set and all players receive the same payoff for being in the same situation, we can simplify our payoff table.  

\noindent Let $N$ be a matrix, where each row $N_i$ contains a discrete distribution of $p$ players over $k$ strategies. The matrix yields $\binom{p+k-1}{p}$ rows. Each distribution over strategies can be simulated (or derived from data), returning a vector of expected rewards $u(N_i)$. Let $U$ be a matrix which captures the rewards corresponding to the rows in $N$, i.e., $U_i = u(N_i)$. We refer to a meta payoff table as $M = (N, U)$. 

\noindent So each row yields a \emph{discrete profile} $(n_{\pi_1}, \ldots, n_{\pi_k})$ indicating exactly how many players play each strategy, with $\sum_j n_{\pi_j}=p$. A strategy profile $\mathbf{x}$ then equals $(\frac{n_{\pi_1}}{p}, \ldots, \frac{n_{\pi_k}}{p})$. 

\noindent Suppose we have a meta-game with $3$ meta-strategies ($|S|=3$) and $6$ players ($|A|=6$) that interact in a $6$-type, this leads to a meta game payoff table of $28$ entries (which is a good reduction from $3^6 cells$. An important advantage of this type of table is that it easily extends to many agents, as opposed to the classical payoff matrix. Table \ref{table:hpt} provides an example for three strategies $\pi_1, \pi_2$ and $\pi_3$. The left-hand side expresses the discrete profiles and corresponds to matrix $N$, while the right-hand side gives the payoffs for playing any of the strategies given the discrete profile and corresponds to matrix $U$. 



\begin{table}[!ht]
\footnotesize
		\begin{center}
		$P = \left( \begin{array}{ccccccc}
		N_{i1}& N_{i2} & N_{i3} & \vline & U_{i1} & U_{i2} & U_{i3} \\ 
		\hline
		6 & 0 & 0 & \vline & 0 & 0 & 0 \\
		& ... & & \vline & & ... & \\
		4 & 0 & 2 & \vline & -0.5 & 0 & 1 \\
		& ... & & \vline & & ... & \\
		0 & 0 & 6 & \vline & 0 & 0 & 0 \\
		\end{array} \right)$ 
		\end{center}
    \caption{\small An example of a meta game payoff table}
    \label{table:hpt}
    \vspace{-1cm}
\end{table}


In order to analyse the evolutionary dynamics of high-level meta-strategies, we also need to estimate the expected payoff of such strategies relative to each other. In evolutionary game theoretic terms, this is the relative fitness of the various strategies, dependent on the current frequencies of those strategies in the population.

In order to approximate the payoff for an arbitrary mix of strategies in an infinite population of replicators distributed over the species according to $\mathbf{x}$, $p$ individuals are drawn randomly from the infinite distribution. The probability for selecting a specific row $N_i$ can be computed from $\mathbf{x}$ and $N_i$ as
\begin{equation*}
  \phantom{.}P(N_i | \mathbf{x}) = \binom{p}{N_{i1}, N_{i2}, \ldots, N_{ik}} \prod_{j=1}^{k} x_j^{N_{ij}}.
\end{equation*}
The expected payoff of strategy $\pi^i$, $r^i(\mathbf{x})$, is then computed as the weighted combination of the payoffs given in all rows:
\begin{equation*}
  \phantom{.}r^i(\mathbf{x}) = \frac{\sum_j P(N_j | \mathbf{x}) U_{ji}}{1 - (1-x_i)^p}.
\end{equation*}
This expected payoff function can be used in Equation~\ref{eq:singleRD} to compute the evolutionary population change according to the replicator dynamics by replacing $(A x)_i$ by $r^i(\mathbf{x})$. Note that we need  to re-normalize (denominator) by ignoring rows that do not contribute to the payoff of a strategy because it is not present in the distribution $N_j$ in the HPT.

\vspace{-\secspace cm}

\subsection{Asymmetric Meta Games}
\noindent One can now wonder how the previously introduced method extends to asymmetric games, which has not been considered in the literature. 
An example of an asymmetric game is the famous battle of the sexes game illustrated in Table \ref{fig:BoS}. In this game both players do have the same strategy sets, i.e., go to the opera or go to the movies, however, the corresponding payoffs for each are different, expressing the differences in preferences that both players have.

\begin{table}[htb]
\vspace{-0.5cm}
\footnotesize
	\centering
\begin{minipage}{.20\textwidth}
   \begin{game}{2}{2}[][]
   	    &  O     &  M    \\
   	 O  &    $3,2$      & $0,0$ \\
   	 M &  $0,0$ & $2,3$ \\
   \end{game}
  \caption{\footnotesize Battle of the Sexes game: strategies $O$ and $M$ correspond to going to the Opera and going to the Movies respectively.}\label{fig:BoS}
\end{minipage}
\begin{minipage}{.30\textwidth}
   \begin{game}{3}{3}[][]
   	  &  $C_1$     &  $C_2$   & $C_3$  \\
   	 $R_1$  &    $r_{11},c_{11}$      & $r_{12},c_{12}$ & $r_{13},c_{13}$\\
   	 $R_2$ &  $r_{21},c_{21}$ & $r_{22},c_{22}$ & $r_{23},c_{23}$\\
   	 $R_3$ & $r_{31},c_{31}$ & $r_{32},c_{32}$ & $r_{33},c_{33}$\\
   \end{game}
   \caption{\footnotesize General 3x3 normal form game.}
   \label{fig:gentab}
\end{minipage}
\vspace{-0.8cm}
\end{table}

\noindent If we aim to carry out a similar evolutionary analysis as in the symmetric case, restricting ourselves to two populations or roles, we will need two meta game payoff tables, one for each player over its own strategy set. We will also need to use the asymmetric version of the replicator dynamics as shown in Equation \ref{eq:asymRD2}. 
Additionally, in order to compute the right payoffs for every situation we will have to interpret a discrete strategy profile in the meta-table slightly different.  Suppose we have a 2-type meta game, with three strategies in each player's strategy set. We introduce a generalisation of our meta-table for both players by means of an example shown in Table \ref{table:asym-hpt}, which corresponds to the general NFG shown in Table \ref{fig:gentab}.

\begin{table}[!ht]
\vspace{-0.3cm}
\footnotesize
		\begin{center}
		$P = \left( \begin{array}{ccccccc}
		N_{i1,j1}& N_{i2,j2} & N_{i3,j3} & \vline & U_{i1,j1} & U_{i2,j2} & U_{i3,j3} \\ 
		\hline
		(1,1) & 0 & 0 & \vline & (r_{11},c_{11}) & 0 & 0 \\
		& ... & & \vline & & ... & \\
		(1,0) & (0,1) & 0 & \vline & (r_{12},0) & (0,c_{12}) & 0 \\
		(0,1) & (1,0) & 0 & \vline & (0,c_{21}) & (r_{21},0) & 0 \\
		& ... & & \vline & & ... & \\
		0 & 0 & (1,1) & \vline & 0 & 0 & (r_{33},c_{33}) \\
		\end{array} \right)$ 
		\end{center}
    \caption{\small An example of an asymmetric meta game payoff table}
    \label{table:asym-hpt}
    \vspace{-1cm}
\end{table}

\noindent Let's have a look at the first entry in Table \ref{table:asym-hpt}, i.e., $[(1,1), 0, 0]$. This entry means that both agents ($i$ and $j$) are playing their first strategy, expressed by $N_{i1,j1}$, meaning the number of agents $N_{i1}$ playing strategy $\pi^1_i$ in the first population equals $1$ and that the number of agents $N_{j1}$ playing strategy $\pi^2_j$ in the second population equals $1$ as well. The corresponding payoff for each player $U_{i1,j1}$ equals $(r_{11},c_{11})$. Now lets have a look at the discrete profiles: $[(1,0),(0,1),0]$ and $[(0,1),(1,0),0]$. The first one means that the first player is playing its first strategy while the second player is playing their second strategy. The corresponding payoffs are $r_{12}$ for the first player and  $c_{12}$ for the second player. 
The profile $[(0,1),(1,0),0]$ shows the reverted situation in which the second player plays his first strategy and the first player plays his second strategy, yielding payoffs $r_{21}$ and $c_{21}$ for the first player and second player respectively. 
In order to turn the table into a similar format as for the symmetric case, we can now introduce $p$ meta-tables, one for each player. More precisely, we get Tables \ref{table:asym-hpt-decomp1} and \ref{table:asym-hpt-decomp2} for players 1 and 2 respectively.

\begin{table}[!ht]
\vspace{-0.5cm}
\footnotesize
		\begin{center}
		$P = \left( \begin{array}{ccccccc}
		N_{i1,j1}& N_{i2,j2} & N_{i3,j3} & \vline & U_{i1,j1} & U_{i2,j2} & U_{i3,j3} \\ 
		\hline
		2 & 0 & 0 & \vline & r_{11} & 0 & 0 \\
		& ... & & \vline & & ... & \\
		1 & 1 & 0 & \vline & r_{12} & r_{21} & 0 \\
		& ... & & \vline & & ... & \\
		0 & 0 & 2 & \vline & 0 & 0 & r_{33} \\
		\end{array} \right)$ 
		\end{center}
    \caption{\small A decomposed asymmetric meta payoff table for Player 1.}
    \label{table:asym-hpt-decomp1}
    \vspace{-1.1cm}
\end{table}

\begin{table}[!ht]
\footnotesize
		\begin{center}
		$Q = \left( \begin{array}{ccccccc}
		N_{i1,j1}& N_{i2,j2} & N_{i3,j3} & \vline & U_{i1,j1} & U_{i2,j2} & U_{i3,j3} \\ 
		\hline
		2 & 0 & 0 & \vline & c_{11} & 0 & 0 \\
		& ... & & \vline & & ... & \\
		1 & 1 & 0 & \vline & c_{12} & c_{21} & 0 \\
		& ... & & \vline & & ... & \\
		0 & 0 & 2 & \vline & 0 & 0 & c_{33} \\
		\end{array} \right)$ 
		\end{center}
    \caption{\small A decomposed asymmetric meta payoff table for Player 2.}
    \label{table:asym-hpt-decomp2}
    \vspace{-0.8cm}
\end{table}

\noindent One needs to take care in correctly interpreting these tables. Let's have a look at row $[1, 1, 0]$ for instance. This should now be interpreted in two ways: one, the first player plays his first strategy while the other player plays his second strategy and he receives a payoff of $r_{12}$, two, the first player plays his second strategy while the other player plays his first strategy and receives a payoff of $r_{21}$.
The expected payoff $r^i(\mathbf{x})$ can now be estimated in the same way as explained for the symmetric case as we will be relying on symmetric replicator dynamics by decoupling asymmetric games in their \textit{symmetric counterparts} (explained in the next section).

\vspace{-\secspace cm}

\subsection{Linking symmetric and asymmetric games}\label{sec:theor}

Here we summarize the most important results on the link between an asymmetric game and its symmetric counterpart games. For a full treatment and discussion of these results see \cite{TuylsSym}. 
In a nutshell, this work proves that if $x,y$ is a Nash equilibrium of the bimatrix game $(A,B)$ (where $x$ and $y$ have the same support\footnote{$x$ and $y$ have the same support if $I_x=I_y$ where $I_x = \{i \; | \; x_i>0\}$ and  $I_y = \{i \; | \; y_i>0\}$}), then $y$ is a Nash equilibrium of the single population, or symmetric, game $A$ and $x$ is a Nash equilibrium of the single population, or symmetric, game $B^\top$. Both symmetric games are called the \emph{counterpart games} of the asymmetric game $(A,B)$. The reverse is also true: If $y$ is a Nash equilibrium of the single population game $A$ and $x$ is a Nash equilibrium of the single population game $B^\top$ (and if $x$ and $y$ have the same support), then $x,y$ is a Nash equilibrium of the game $(A,B)$.
In our empirical analysis, we use this property to analyze an asymmetric games $(A,B)$ by looking at the counterpart single population games $A$ and $B^\top$.






\vspace{-\secspace cm}
\section{Theoretical Insights}

\label{TheoreticalInsights}

As illustrated in the previous section the procedure for empirical meta-game analysis consists of two parts. Firstly, one needs to construct an empirical meta-game utility function for each player. This step can be performed using logs of interactions between players, or by playing the game sufficiently enough. Secondly, one expects that analyzing the empirical game will give insights in the true underlying game itself (i.e. the game from which we sample).
This section provides insights in the following: how much data is enough to generate a good approximation of the true underlying game? Is uniform sampling over actions or strategies the right method? 

\vspace{-\secspace cm}

\subsection{Main Lemma}

Sometimes players receive a stochastic reward $R^i(\pi^1, \dots, \pi^p)$ for a given joint action $\bm{\pi}$. The underlying game we study is $r^i(\pi^1, \dots,\pi^p) = E \left[ R^i(\pi^1, \dots,\pi^p) \right]$ and for the sake of simplicity the joint action of every player but player $i$ will be written $\bm{\pi}^{-i}$.
In the two following definitions, we introduce the concept of Nash equilibrium and $\epsilon$-Nash equilibrium in $p$-player games (as we only introduced it in the $2$-player case):

\textbf{Definition :} A joint strategy $\bm{x} = (x^1, \dots,x^p) = (x^{-i}, \bm{x}^{-i})$ is a Nash equilibrium if for all $i$:
$$E_{\bm{\pi} \sim \bm{x}} \left[ r^i(\bm{\pi})\right] = \max_{\pi^i} E_{\bm{\pi}^{-i} \sim \bm{x}^{-i}} \left[ r^i(\pi^i, \bm{\pi}^{-i})\right]$$

\textbf{Definition :} A joint strategy $\bm{x} = (x^1, \dots,x^p) = (x^{-i}, \bm{x}^{-i})$ is an $\epsilon$-Nash equilibrium if for all $i$:
$$\max_{\pi^i} E_{\bm{\pi}^{-i} \sim \bm{x}^{-i}} \left[ r^i(\pi^i, \bm{\pi}^{-i})\right] - E_{\bm{\pi} \sim \bm{x}} \left[ r^i(\bm{\pi})\right] \leq \epsilon$$

When running an analysis on a meta game, we do not have access to the average reward function $r^i(\pi^1, \dots,\pi^p)$ but to an empirical estimate $\hat{r}^i(\pi^1, \dots,\pi^p)$. The following lemma shows that a Nash equilibrium for the empirical game $\hat{r}^i(\pi^1, \dots,\pi^p)$ is an $2\epsilon$-Nash equilibrium for the game $r^i(\pi^1, \dots,\pi^p)$ where $\epsilon = \sup_{\bm{\pi},i} |\hat{r}^i(\bm{\pi})-r^i(\bm{\pi})|$.

\textbf{Lemma:} If $\bm{x}$ is a Nash equilibrium for $\hat{r}^i(\pi^1, \dots,\pi^p)$, then it is an $2\epsilon$-Nash equilibrium for the game $r^i(\pi^1, \dots,\pi^p)$ where $\epsilon = \sup_{\bm{\pi},i} |r^i(\bm{\pi})-\hat{r}^i(\bm{\pi})|$.

\textbf{Proof:}\newline
First we have the following relation:
\begin{align*}
    &E_{\bm{\pi} \sim \bm{x}} \left[ r^i(\bm{\pi})\right] = E_{\bm{\pi} \sim \bm{x}} \left[ \hat{r}^i(\bm{\pi})\right] + E_{\bm{\pi} \sim \bm{x}} \left[ r^i(\bm{\pi}) - \hat{r}^i(\bm{\pi})\right]
\end{align*}
Then:
\footnotesize
\begin{align*}
    E_{\bm{\pi}^{-i} \sim \bm{x}^{-i}} \left[ r^i(\pi^i, \bm{\pi}^{-i})\right] &= E_{\bm{\pi}^{-i} \sim \bm{x}^{-i}} \left[ \hat{r}^i(\pi^i, \bm{\pi}^{-i})\right]\\
    &\quad + E_{\bm{\pi}^{-i} \sim \bm{x}^{-i}} \left[ r^i(\pi^i, \bm{\pi}^{-i}) - \hat{r}^i(\pi^i, \bm{\pi}^{-i})\right]\\
    \max_{\pi^i} E_{\bm{\pi}^{-i} \sim \bm{x}^{-i}} \left[ r^i(\pi^i, \bm{\pi}^{-i})\right] &\leq \max_{\pi^i} E_{\bm{\pi}^{-i} \sim \bm{x}^{-i}} \left[ \hat{r}^i(\pi^i, \bm{\pi}^{-i})\right]\\
    & \quad + \max_{\pi^i} E_{\bm{\pi}^{-i} \sim \bm{x}^{-i}} \left[ r^i(\pi^i, \bm{\pi}^{-i}) - \hat{r}^i(\pi^i, \bm{\pi}^{-i})\right]
\end{align*}
\normalsize
Finally,
\footnotesize
\begin{align*}
    &\max_{\pi^i} E_{\bm{\pi}^{-i} \sim \bm{x}^{-i}} \left[ r^i(\pi^i, \bm{\pi}^{-i})\right] - E_{\bm{\pi} \sim \bm{x}} \left[ r^i(\bm{\pi})\right]\\
    &\qquad\qquad\qquad \leq \underbrace{\max_{\pi^i} E_{\bm{\pi}^{-i} \sim \bm{x}^{-i}} \left[ \hat{r}^i(\pi^i, \bm{\pi}^{-i})\right] - E_{\bm{\pi} \sim \bm{x}} \left[ \hat{r}^i(\bm{\pi})\right]}_{=0 \textrm{ since $\bm{x}$ is a Nash equilibrium for $\hat{r}^i$}}\\
    &\qquad\qquad\qquad\qquad \underbrace{ + \max_{\pi^i} E_{\bm{\pi}^{-i} \sim \bm{x}^{-i}} \left[ r^i(\pi^i, \bm{\pi}^{-i}) - \hat{r}^i(\pi^i, \bm{\pi}^{-i})\right]}_{\leq \epsilon}\\
    &\qquad\qquad\qquad\qquad \underbrace{ - E_{\bm{\pi} \sim \bm{x}} \left[ r^i(\bm{\pi}) - \hat{r}^i(\bm{\pi})\right]}_{\leq \epsilon}
\end{align*}
\normalsize
\qed

This lemma shows that if one can control  the difference between $|r^i(\bm{\pi})-\hat{r}^i(\bm{\pi})|$ uniformly over players and actions, then an equilibrium for the empirical game $\hat{r}^i(\pi^1, \dots,\pi^p)$ is almost an equilibrium for the game defined by the average reward function $r^i(\pi^1, \dots,\pi^p)$.

\subsection{Finite Samples Analysis}
This section details some concentration results. In practice, we often have access to a batch of observations of the underlying game. We will run our analysis on an empirical estimate of the game denoted by $\hat{r}^i(\bm{\pi})$. The question then will be either with which confidence can we say that a Nash equilibrium for $\bm{\hat{r}}$ is a $2\epsilon$-Nash equilibrium, or for a fixed confidence, for which $\epsilon$ can we say that a Nash equilibrium for $\bm{\hat{r}}$ is a $2\epsilon$-Nash equilibrium for $\bm{r}$. In the case we have access to game play, the question is how many samples $n$ do we need to assess that a Nash equilibrium for $\bm{\hat{r}}$ is a $2\epsilon$-Nash equilibrium for $\bm{r}$ for a fixed confidence and a fixed $\epsilon$.
For the sake of simplicity, we will assume that all payoff are bounded in $[0, 1]$.

\subsubsection{The batch scenario}
\label{batchScenario}

Here we assume that we are given $n(i,\bm{\pi})$ independent samples to compute the empirical average $\hat{r}^i(\bm{\pi})$. Then, by applying Hoeffding’s inequality we can prove the following result:
\footnotesize
$$P\left( \sup_{\bm{\pi},i} |r^i(\bm{\pi})-\hat{r}^i(\bm{\pi})| < \epsilon \right) \geq \prod_{i \in \{1,\dots,p\}} \prod_{\bm{\pi}} \left(1-2e^{\left(-2\epsilon^2 n(i,\bm{\pi})\right)}\right)$$
\normalsize
\subsubsection{uniform sampling}
In this section we assume that we have a budget of $n$ samples per joint actions $\bm{\pi}$ and per player $i$. In that case we have the following bound:
\footnotesize
$$P\left( \sup_{\bm{\pi},i} |r^i(\bm{\pi})-\hat{r}^i(\bm{\pi})| < \epsilon \right) \geq \left(1-2e^{\left(-2\epsilon^2 n\right)}\right)^{|S^1| \times \dots \times |S^p| \times p}$$
\normalsize
Then, If we want $\sup_{\bm{\pi},i} |r^i(\bm{\pi})-\hat{r}^i(\bm{\pi})| < \epsilon $ with a probability of at least $1-\delta$ we need at least $n= - \frac{\ln\left(1-(1-\delta)^{\frac{1}{|S^1| \times \dots \times |S^p| \times p}}\right)}{2\epsilon^2}$



\vspace{-\secspace cm}
\section{Experiments}\label{sec:exps}

This section presents experiments that illustrate the meta-game approach and its feasibility for examining strengths and weaknesses of higher-level strategies in various domains, including \textit{AlphaGo}, Colonel Blotto, and the meta-game generated by PSRO. Note that we restrict the meta-games to three strategies, as we can nicely visualise this in a phase plot, and these still provide useful information about the dynamics in the full strategy spaces.


\vspace{-\secspace cm}

\subsection{AlphaGo}

The data set under study consists of $7$ \textit{AlphaGo} variations and a a number of different Go strategies such as Crazystone and Zen (previously the state-of-the-art). $\alpha$ stands for the algorithm and the indexes $r,v,p$ for the use of respectively \emph{rollouts}, \emph{value nets} and \emph{policy nets} (e.g. $\alpha_{rvp}$ uses all 3). For a detailed description of these strategies see \cite{DSilverHMGSDSAPL16}. 
The meta-game under study here concerns a $2$-type NFG with $|S|=9$. We will look at various $2$-faces of the larger simplex. Table 9 in \cite{DSilverHMGSDSAPL16} summarises all wins and losses between these various strategies (meeting several times), from which we can compute meta-game payoff tables.

\subsubsection{Experiment 1: strong strategies}

\noindent This first experiment examines three of the strongest \textit{AlphaGo} strategies in the data-set, i.e., $\alpha_{rvp}, \alpha_{vp}, \alpha_{rp}$.
\noindent As a first step we created a meta-game payoff table involving these three strategies, by looking at their pairwise interactions in the data set (summarised in Table 9 of \cite{DSilverHMGSDSAPL16}). This set contains data for all strategies on how they interacted with the other $8$ strategies, listing the win rates that strategies achieved against one another (playing either as white or black) over several games.
The meta-game payoff table derived for these three strategies is described in Table \ref{table:exp1Go}. 

\begin{table}[!ht]
\vspace{-0.3cm}
\footnotesize
		\begin{center}
		$\left( \begin{array}{ccccccc}
		\alpha_{rvp} & \alpha_{vp} & \alpha_{rp} & \vline & U_{i1} & U_{i2} & U_{i3} \\ 
		\hline
		2 & 0 & 0 & \vline & 0.5 & 0 & 0 \\
		1 & 0 & 1 & \vline & 0.95 & 0 & 0.05 \\
		0 & 2 & 0 & \vline & 0 & 0.5 & 0 \\
		1 & 1 & 0 & \vline & 0.99 & 0.01 & 0 \\
		0 & 0 & 2 & \vline & 0 & 0 & 0.5 \\
		0 & 1 & 1 & \vline & 0 & 0.39 & 0.61 \\
		\end{array} \right)$ 
		\end{center}
    \caption{\small Meta-game payoff table generated from Table 9 in \cite{DSilverHMGSDSAPL16} for strategies $\alpha_{rvp}, \alpha_{vp}, \alpha_{rp}$}
    \label{table:exp1Go}
    \vspace{-0.8cm}
\end{table}

\begin{figure*}[!tbp]
  \centering
  \begin{minipage}[b]{0.33\textwidth}
     \includegraphics[width=\textwidth]{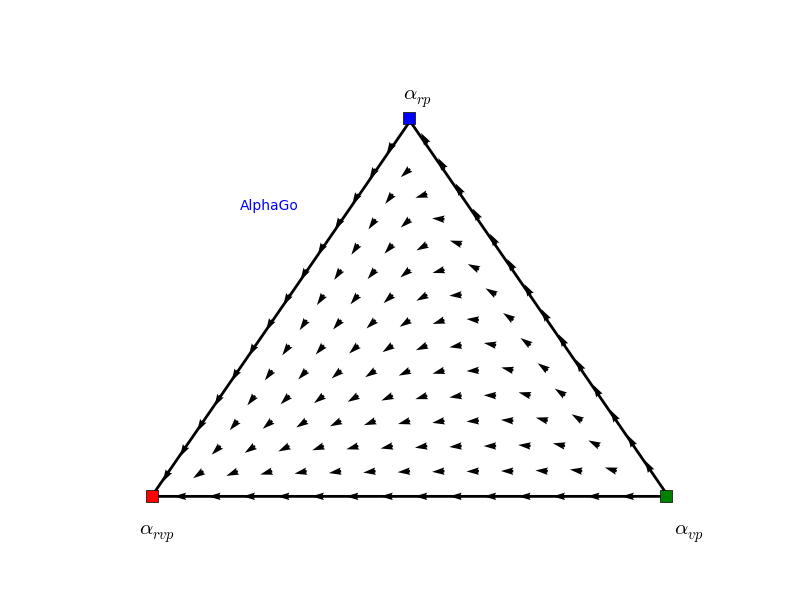}
  \vspace{-1.2cm}
  \caption{\small Directional field plot for the 2-face consisting of strategies $\alpha_{rvp}, \alpha_{vp}, \alpha_{rp}$}
    \label{fig:alphago-exp1-df}
  \end{minipage}
 \qquad
  \begin{minipage}[b]{0.33\textwidth}
\includegraphics[width=\textwidth]{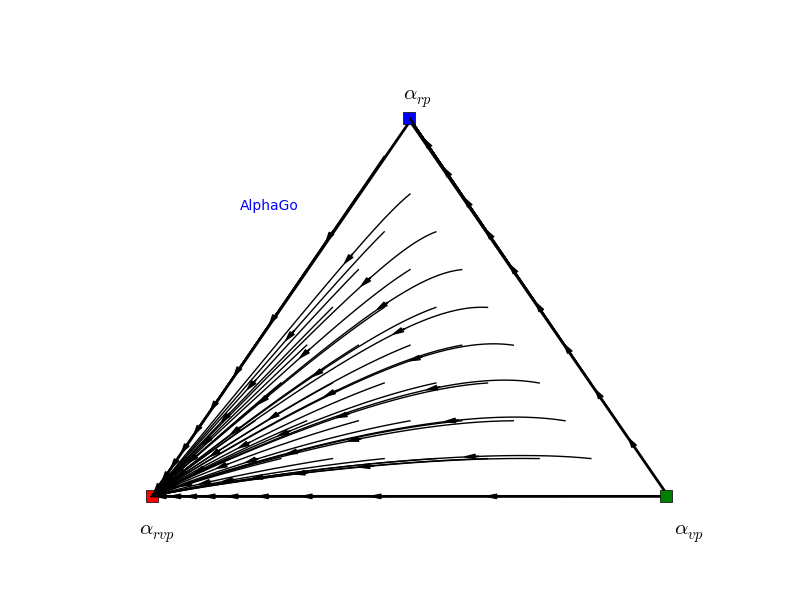}
  \vspace{-1.2cm}
  \caption{\small Trajectory plot for the 2-face consisting of strategies $\alpha_{rvp}, \alpha_{vp}, \alpha_{rp}$}
    \label{fig:alphago-exp1-tp}
  \end{minipage}
\end{figure*}

\noindent In Figure \ref{fig:alphago-exp1-df} we have plotted the directional field of the meta-game payoff table using the replicator dynamics  for a number of strategy profiles $\mathbf{x}$ in the simplex strategy space. From each of these points in strategy space an arrow indicates the direction of flow, or change, of the population composition over the three strategies. Figure \ref{fig:alphago-exp1-tp} shows a corresponding trajectory plot. From these plots one can easily observe that strategy $\alpha_{rvp}$ is a strong attractor and consumes the entire strategy space over the three strategies. This restpoint is also a Nash equilibrium. This result is in line with what we would expect from the knowledge we have of the strengths of these various learned policies. Still, the arrows indicate how the strategy landscape flows into this attractor and therefore provides useful information as we will discuss later.

\subsubsection{Experiment 2: evolution and transitivity of strengths}

\begin{figure*}[!tbp]
\vspace{-0.3cm}
\centering
\begin{minipage}{0.33\textwidth}
    \begin{subfigure}{\textwidth}
    \centering
    \vspace{-0.1cm}
    \includegraphics[width=\textwidth]{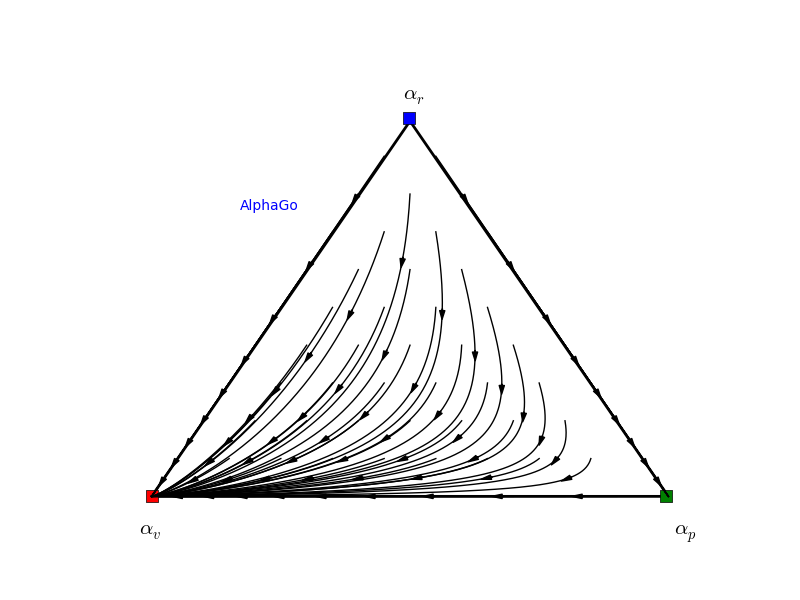}
    \vspace{-1cm}
    \caption{\footnotesize Trajectory plot for $\alpha_v$, $\alpha_p$, and $\alpha_r$}
    \end{subfigure}\\
    \begin{subfigure}{\textwidth}
    \centering
    \includegraphics[width=\textwidth]{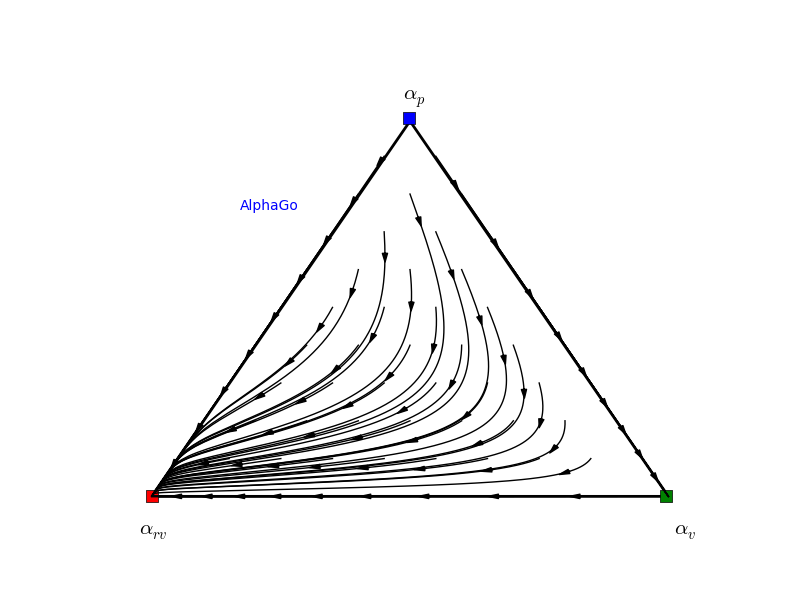}
    \vspace{-1cm}
    \caption{\footnotesize Trajectory plot for $\alpha_{rv}$, $\alpha_v$, and $\alpha_p$}
    \end{subfigure}%
    \vspace{-0.3cm}
    \caption{}\label{AG:exp2}
\end{minipage}%
\hfill
\begin{minipage}{.33\textwidth}
    \begin{subfigure}{\textwidth}
    \centering
    \vspace{-0.1cm}
    \includegraphics[width=\textwidth]{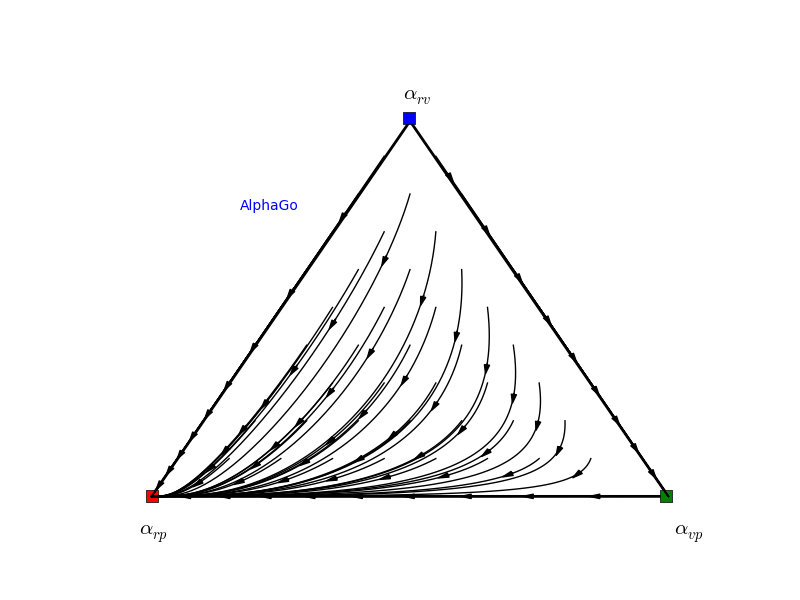}
    \vspace{-1cm}
    \caption{\footnotesize Trajectory plot for $\alpha_{rp}$, $\alpha_{vp}$, and $\alpha_{rv}$}
    \end{subfigure}\\
    \begin{subfigure}{\textwidth}
    \centering
    \includegraphics[width=\textwidth]{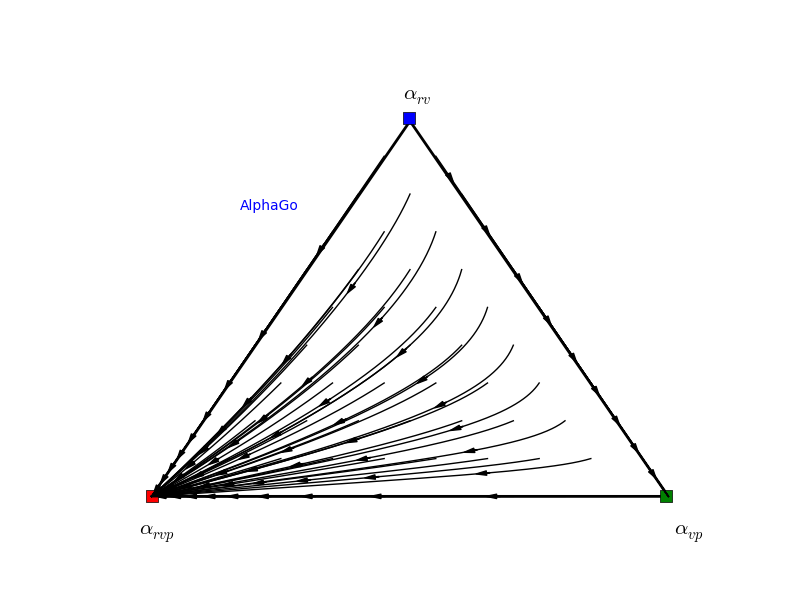}
    \vspace{-1cm}
    \caption{\footnotesize Trajectory plot for $\alpha_{rvp}$, $\alpha_{vp}$, and $\alpha_{rv}$}
    \end{subfigure}%
    \vspace{-0.3cm}
    \caption{}\label{AG:exp3}
\end{minipage}
\hfill
\begin{minipage}{.33\textwidth}
    \begin{subfigure}{\textwidth}
    \centering
    \includegraphics[width=\textwidth]{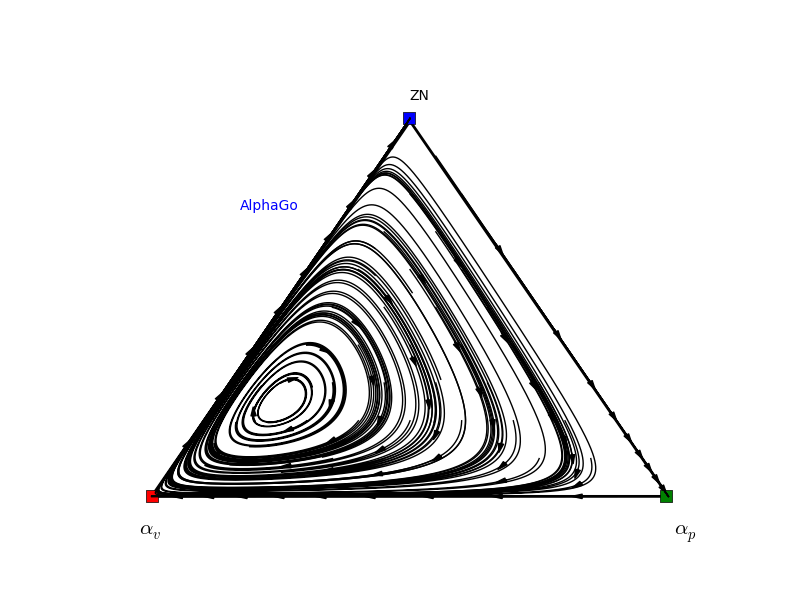}
    \vspace{-1.2cm}
    \caption{}
    \end{subfigure}%
    \vspace{-0.3cm}
    \caption{\footnotesize Intransitive behaviour for $\alpha_v$, $\alpha_p$, and $Zen$.}\label{AG:exp4}
\end{minipage}
\end{figure*}

\noindent We start by investigating the 2-face simplex involving strategies $\alpha_{rp}$, $\alpha_{vp}$ and $\alpha_{rv}$, for which we created a meta-game payoff table similarly as in the previous experiment (not shown). The evolutionary dynamics of this 2-face can be observed in Figure \ref{AG:exp3}a. Clearly strategy $\alpha_{rp}$ is a strong attractor and dominates the two other strategies. We now replace this attractor by strategy $\alpha_{rvp}$ and plot its evolutionary dynamics in Figure \ref{AG:exp3}b.
What can be observed from both trajectory plots in Figure \ref{AG:exp3} is that the curvature is less pronounced in plot \ref{AG:exp3}b than it is in plot \ref{AG:exp3}a. The reason for this is that the difference in strength between $\alpha_{rv}$ and $\alpha_{vp}$ is less obvious in the presence of an even stronger attractor than $\alpha_{rp}$. This means that $\alpha_{rvp}$ is now pulling much stronger on both $\alpha_{rv}$ and $\alpha_{vp}$ and consequently the flow goes more directly to $\alpha_{rvp}$. 
So even when a strategy space is dominated by one strategy, the curvature (or curl) is a promising measure for the strength of a meta-strategy.

What is worthwhile to observe from the \textit{AlphaGo} dataset, and illustrated as a series in Figures \ref{AG:exp2} and \ref{AG:exp3}, is that there is clearly an incremental increase in the strength of the \textit{AlphaGo} algorithm going from version $\alpha_r$ to $\alpha_{rvp}$,  building on previous strengths, without any intransitive behaviour occurring, when only considering a strategy space formed by the \textit{AlphaGo} versions. 

Finally, as discussed in Section~\ref{TheoreticalInsights}, we can now examine how good of an approximation an estimated game is. In the \textit{AlphaGo} domain we only do this analysis for the games displayed in Figures \ref{AG:exp3}a and \ref{AG:exp3}b, as it is similar for the other experiments. We know that \emph{$\alpha_{rp}$} is a Nash equilibrium of the estimated game analyzed in Figure~\ref{AG:exp3}a (meta Table not shown). The outcome of $\alpha_{rp}$ against $\alpha_{rv}$ was estimated with $n_{\alpha_{rp},\alpha_{rv}} = 63$ games (for the other pair of strategies we have $n_{\alpha_{vp},\alpha_{rp}} = 65$ and $n_{\alpha_{vp},\alpha_{rv}} = 133$). Because of the symmetry of the problem, the bound in section~\ref{batchScenario} is reduced to:

\vspace{-0.2cm}
\footnotesize
\begin{align*}
P\left( \sup_{\bm{\pi},i} |r^i(\bm{\pi})-\hat{r}^i(\bm{\pi})| < \epsilon \right) \geq  \left(1-2e^{\left(-2\epsilon^2 n_{\alpha_{rp},\alpha_{rv}}\right)}\right) &\times \left(1-2e^{\left(-2\epsilon^2 n_{\alpha_{vp},\alpha_{rp}}\right)}\right) \\
&\times \left(1-2e^{\left(-2\epsilon^2 n_{\alpha_{vp},\alpha_{rv}}\right)}\right)
\end{align*}
\normalsize

Therefore, we can conclude that the strategy \emph{$\alpha_{rp}$} is an $2\epsilon$-Nash equilibrium (with $\epsilon=0.15$) for the real game with probability at least $0.78$. The same calculation would also give a confidence of $0.85$ for the RD studied in Figure~\ref{AG:exp3}b for an $\epsilon = 0.15$ (as the number of samples are $(n_{\alpha_{rv},\alpha_{vp}},n_{\alpha_{vp},\alpha_{rvp}},n_{\alpha_{rvp},\alpha_{rv}}) = (65,106,91)$).

\subsubsection{Experiment 3: cyclic behaviour}
A final experiment investigates what happens if we add a \textit{pre-AlphaGo} state-of-the-art algorithm to the strategy space. We have observed that even though $\alpha_{rvp}$ remains the strongest strategy, dominating all other \textit{AlphaGo} versions and previous state-of-the-art algorithms, cyclic behaviour can occur, something that cannot be measured or seen from Elo ratings.\footnote{An Elo rating or score is a measure to express the relative strength of a player, or strategy. It was named after Arpad Elo and originally introduced to rate chess players. For an introduction see e.g. \cite{Coulom08}} More precisely, we constructed a meta-game payoff table for strategies $\alpha_v$, $\alpha_p$ and $Zen$ (one of the previous commercial state-of-the-art algorithms). In Figure \ref{AG:exp4} we have plotted the evolutionary dynamics for this meta-game, and as can be observed there is a mixed equilibrium in strategy space, around which the dynamics cycle, indicating that $Zen$ is capable of introducing in-transitivity, as $\alpha_v$ dominates $\alpha_p$, $\alpha_p$ dominates $Zen$ and $Zen$ dominates $\alpha_v$.

\vspace{-\secspace cm}

\subsection{Colonel Blotto}

\noindent Colonel Blotto is a resource allocation game originally introduced by Borel \cite{Borel}. Two players interact, each allocating $m$ troops over $n$ locations. They do this separately without communication, after which both distributions are compared to determine the winner. When a player has more troops in a specific location, it wins that location. The player winning the most locations wins the game. This game has many game theoretic intricacies, for an analysis see \cite{KohliKBHSG12}. Kohli et al. have run Colonel Blotto on Facebook (project Waterloo), collecting data describing how humans play this game, with each player having $m=100$ troops and considering $n=5$ battlefields. The number of strategies in the game is vast: a game with $m$ troops and $n$  locations has $\binom{m + n - 1}{n - 1}$ strategies.

Based on Kohli et al. we carry out a meta game analysis of the \emph{strongest strategies} and the\emph{most frequently played strategies} on Facebook. We have a look at several $3$-strategy simplexes, which can be considered as $2$-faces of the entire strategy space.

\noindent An instance of a strategy in the game of Blotto will be denoted as follows: $[t_1,t_2,t_3,t_4,t_5]$ with $\sum_it_i=100$. All permutations $\sigma_i$ in this division of troops belong to the same strategy. 
We assume that permutations are chosen uniformly by a player. Note that in this game there is no need to carry out the theoretical analysis of the approximation of the meta-game, as we are are not examining heuristics or strategies over Blotto strategies, but rather these strategies themselves, for which the payoff against any other strategy will always be the same (by computation). Nevertheless, carrying out a meta-game analysis reveals interesting information.

\subsubsection{Experiment 1: Top performing strategies}

 In this first experiment we examine the dynamics of the simplex consisting of the three best scoring strategies from the study of \cite{KohliKBHSG12}: $[36,35,24,3,2]$, $[37,37,21,3,2]$, and $[35,35,26,2,2]$, see Table \ref{tab:blotto_strategies_strong}. 
 In a first step we compute a meta-game payoff table for these three strategies. The interactions are pairwise, and the expected payoff can be easily computed, assuming a uniform distribution for different permutations of a strategy. This normalised payoff is shown in Table \ref{table:exp1Blotto}.

\begin{table}
\vspace{-1cm}
\footnotesize
\begin{center}
\begin{tabular}{ |p{2cm}|p{1cm}|p{1cm}|  }
 \hline
 \multicolumn{3}{|c|}{Strongest strategies} \\
 \hline
 Strategy & Frequency & Win rate\\
 \hline
 $[36, 35, 24, 3, 2]$   & 1 & .74   \\
 $[37, 37, 21, 3, 2]$ &   17 & .73\\
 $[35, 35, 26, 2, 2]$ & 1 & .73\\
 $[35, 34, 25, 3, 3]$ & 3 & .70\\
 $[35, 35, 24, 3, 3]$ & 13 & .70 \\
 \hline
\end{tabular}
\end{center}
\caption{\small 5 of the strongest strategies played on Facebook.}
\label{tab:blotto_strategies_strong}
\vspace{-1.1cm}
\end{table}

\begin{table}[!ht]
\vspace{-0.2cm}
\footnotesize
		\begin{center}
		$\left( \begin{array}{ccccccc}
		s_1 & s_2 & s_3 & \vline & U_{i1} & U_{i2} & U_{i3} \\ 
		\hline
		2 & 0 & 0 & \vline & 0.5 & 0 & 0 \\
		1 & 0 & 1 & \vline & 0.66 & 0 & 0.34 \\
		0 & 2 & 0 & \vline & 0 & 0.5 & 0 \\
		1 & 1 & 0 & \vline & 0.33 & 0.67 & 0 \\
		0 & 0 & 2 & \vline & 0 & 0 & 0.5 \\
		0 & 1 & 1 & \vline & 0 & 0.75 & 0.25 \\
		\end{array} \right)$ 
		\end{center}
    \caption{\small Meta-game payoff table generated for strategies $s_1=[36,35,24,3,2]$, $s_2=[37,37,21,3,2]$, and $s_3=[35,35,26,2,2]$.}
    \label{table:exp1Blotto}
    \vspace{-0.5cm}
\end{table}

\begin{figure*}[!tbp]
  \centering
  \begin{minipage}[b]{0.33\textwidth}
  \begin{subfigure}{\textwidth}
     \includegraphics[width=\textwidth]{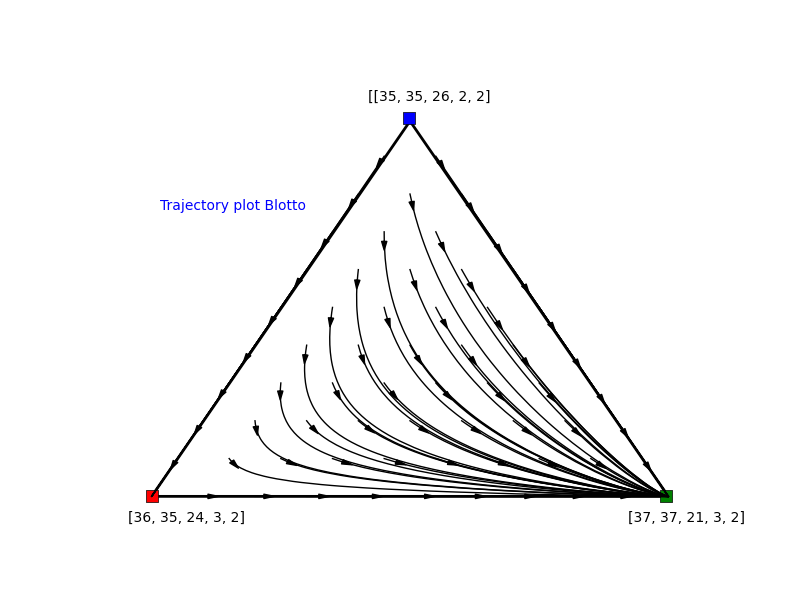}
    \vspace{-1.2cm}
    \caption{}
    \end{subfigure}
  \end{minipage}
  \begin{minipage}[b]{0.33\textwidth}
  \begin{subfigure}{\textwidth}
\includegraphics[width=\textwidth]{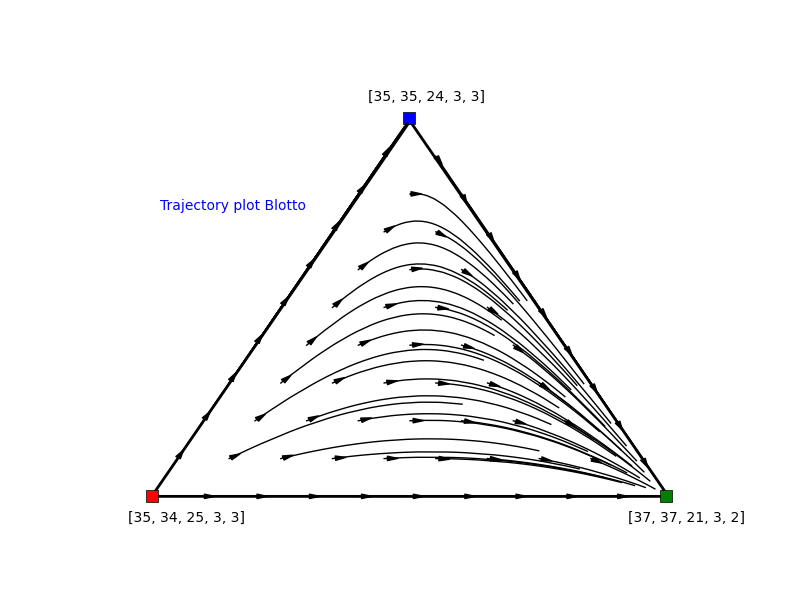}
   \vspace{-1.2cm}
   \caption{}
    \end{subfigure}
  \end{minipage}
  \vspace{-0.5cm}
  \caption{\small (a) dynamics of $[36,35,24,3,2]$, $[37,37,21,3,2]$, and $[35,35,26,2,2]$. (b) dynamics of $[35,34,25,3,3]$, $[37,37,21,3,2]$, and $[35,35,24,3,3]$. }\label{fig:blotto-exps}
\end{figure*}

\noindent Using table \ref{table:exp1Blotto} we can compute evolutionary dynamics using the standard replicator equation. The resulting trajectory plot can be observed in Figure \ref{fig:blotto-exps}a. 
The first thing we see is that we have one strong attractor, i.e, strategy $s_2=[37,37,21,3,2]$ and there is transitive behaviour, meaning that $[36,35,24,3,2]$ dominates $[35,35,26,2,2]$, $[37,37,21,3,2]$ dominates $[36,35,24,3,2]$, and $[37,37,21,3,2]$ dominates $[35,35,26,2,2]$. Although $[37,37,21,3,2]$ is the strongest strategy in this $3$-strategy meta-game, the win rates (computed over all played strategies in project Waterloo) indicate that strategy $[36,35,24,3,2]$ was more successful on Facebook. The differences are minimal, and on average it is better to choose $[37,37,21,3,2]$, which was also the most frequently chosen strategy from the set of strong strategies, see Table \ref{tab:blotto_strategies_strong}.
We show a similar plot for the evolutionary dynamics of strategies $[35,34,25,3,3]$, $[37,37,21,3,2]$, and $[35,35,24,3,3]$ in Figure \ref{fig:blotto-exps}b, which are three of the most frequently played strong strategies from Table \ref{tab:blotto_strategies_strong}. 



\subsubsection{Experiment 2: most frequently played strategies}

\begin{figure*}[!tbp]
\vspace{-0.5cm}
  \centering
  \begin{minipage}[b]{0.33\textwidth}
  \begin{subfigure}{\textwidth}
     \includegraphics[width=\textwidth]{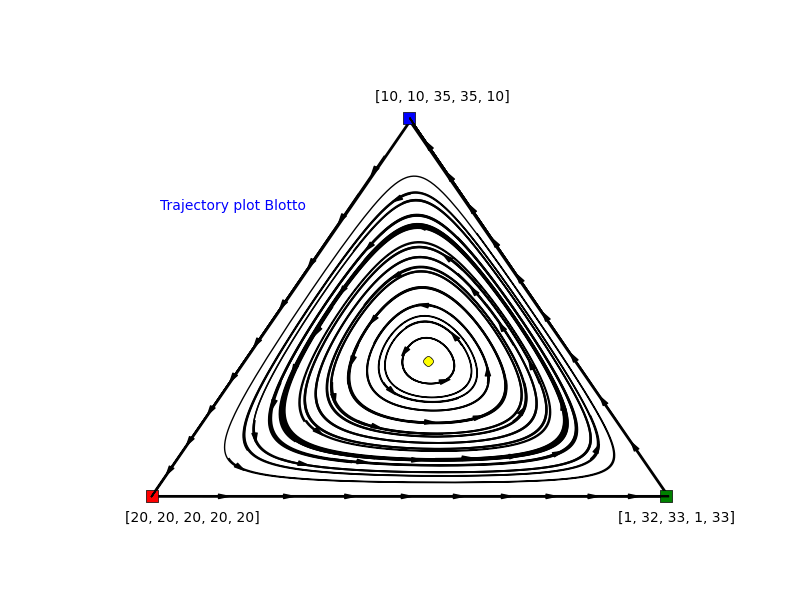}
    \vspace{-1.2cm}
    \caption{}
    \end{subfigure}
  \end{minipage}
  \begin{minipage}[b]{0.33\textwidth}
  \begin{subfigure}{\textwidth}
\includegraphics[width=\textwidth]{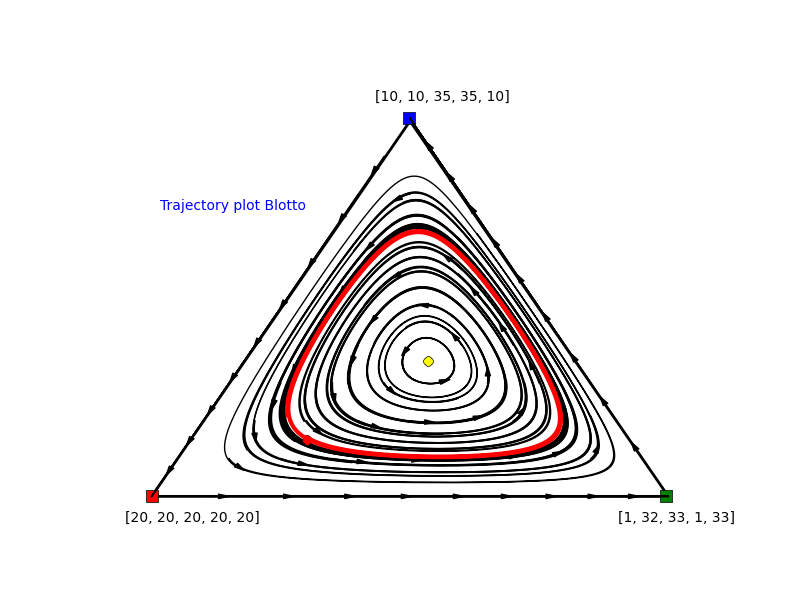}
   \vspace{-1.2cm}
   \caption{}
    \end{subfigure}
  \end{minipage}
    \begin{minipage}[b]{0.33\textwidth}
  \begin{subfigure}{\textwidth}
\includegraphics[width=\textwidth]{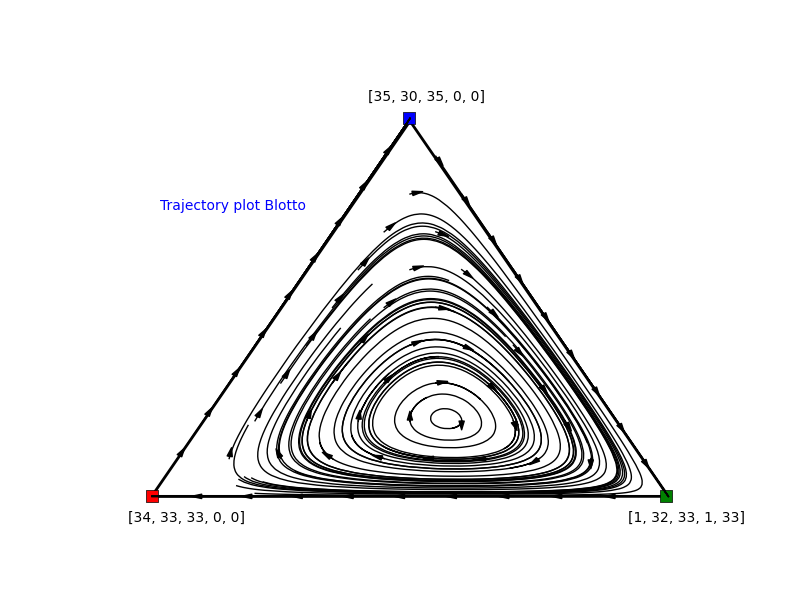}
   \vspace{-1.2cm}
   \caption{}
    \end{subfigure}
  \end{minipage}
  \vspace{-0.5cm}
  \caption{\small Dynamics of 3 $2$-faces of the $8$-simplex: (a) Nash eq. (b) Human play (c) Another example of intransitive behaviour}\label{fig:exp2BlottoNash}
  \vspace{-0.3cm}
\end{figure*}

We compared the evolutionary dynamics of the eight most frequently played strategies and present here a selection of some of the results. The meta-game under study in this domain concerns a 2-type repeated NFG G with $|S|=8$. 
We will look at various $2$-faces of the $8$-simplex.
The top eight most frequently played strategies are shown in Table \ref{tab:blotto_strategies_frequent}.
\begin{table}
\footnotesize
\begin{center}
\begin{tabular}{ |p{2cm}|p{2cm}|  }
 \hline
 \multicolumn{2}{|c|}{Most played strategies} \\
 \hline
 Strategy & Frequency \\
 \hline
 $[34, 33, 33, 0, 0]$   & 271    \\
 $[20, 20, 20, 20, 20]$ &   235 \\
 $[33, 1, 33, 0, 33]$ & 127 \\
 $[1, 32, 33, 1, 33]$ & 97 \\
 $[35, 30, 35, 0, 0]$ & 68 \\
 $[0, 100, 0, 0, 0]$ & 67 \\
 $[10, 10, 35, 35, 10]$ & 58 \\
 $[25, 25, 25, 25, 0]$ & 50 \\
 \hline
\end{tabular}
\end{center}
\caption{\small The 8 most frequently played strategies on Facebook.}
\label{tab:blotto_strategies_frequent}
\vspace{-1cm}
\end{table}
First we investigate the strategies $[20,20,20,20,20]$, $[1,32,33,1,33]$, and $[10,10,35,35,10]$ from our strategy set. In Table \ref{table:exp2Blotto} we show the resulting meta-game payoff table of this $2$-face simplex. Using this table we can again compute the replicator dynamics and investigate the trajectory plots in Figure \ref{fig:exp2BlottoNash}a. We observe that the dynamics cycle around a mixed Nash equilibrium (every interior rest point
is a Nash equilibrium). This intransitive behaviour makes sense by looking at the pairwise interactions between strategies and the corresponding payoffs they receive from Table \ref{table:exp1Blotto}. The expected payoff for $[20,20,20,20,20]$, when playing against $[1,32,33,1,33]$ will be lower than the expected payoff for $[1,32,33,1,33]$. Similarly, $[1,32,33,1,33]$ will be dominated by  $[10,10,35,35,10]$ when they meet, and to make the cycle complete, $[10,10,35,35,10]$ will receive a lower expected payoff against $[20,20,20,20,20]$. As such, the behaviour will cycle around a the Nash equilibrium.

\begin{table}[!h]
\vspace{-0.2cm}
\footnotesize
		\begin{center}
		$\left( \begin{array}{ccccccc}
		s_1 & s_2 & s_3 & \vline & U_{i1} & U_{i2} & U_{i3} \\ 
		\hline
		2 & 0 & 0 & \vline & 0.5 & 0 & 0 \\
		1 & 0 & 1 & \vline & 1 & 0 & 0 \\
		0 & 2 & 0 & \vline & 0 & 0.5 & 0 \\
		1 & 1 & 0 & \vline & 0 & 1 & 0 \\
		0 & 0 & 2 & \vline & 0 & 0 & 0.5 \\
		0 & 1 & 1 & \vline & 0 & 0.1 & 0.9 \\
		\end{array} \right)$ 
		\end{center}
    \caption{\small Meta-game payoff table generated for strategies $s_1=[20,20,20,20,20]$, $s_2=[1,32,33,1,33]$, and $s_3=[10,10,35,35,10]$.} 
    \label{table:exp2Blotto}
    \vspace{-0.7cm}
\end{table}

\noindent An interesting question is where human players are situated in this cyclic behaviour landscape. In Figure \ref{fig:exp2BlottoNash}b we show the same trajectory plot but added a red marker to indicate the strategy profile based on the frequencies of these 3 strategies played by human players. This is derived from Table \ref{tab:blotto_strategies_frequent} and the profile vector is $(0.6,0.25,0.15)$. If we assume that the human agents optimise their behaviour in a \emph{survival of the fittest} style they will cycle along the red trajectory.
In Figure \ref{fig:exp2BlottoNash}c we illustrate similar intransitive behaviour for three other frequently played strategies.

\vspace{-\secspace cm}

\subsection{PSRO-generated Meta-Game}

We now turn our attention to an asymmetric game. Policy Space Response Oracles (PSRO) is a multiagent reinforcement learning process that reduces the strategy space of large extensive-form games via iterative best response computation. PSRO can be seen as a generalized form of fictitious play that produces approximate best responses, with arbitrary distributions over generated responses computed by meta-strategy solvers. 
One application of PSRO was applied to a commonly-used benchmark problem known as Leduc poker~\citep{Southey05}, except with a fixed action space and penalties for taking illegal moves. Therefore PSRO learned to play from scratch, without knowing which moves were legal. Leduc poker has a deck of 6 cards (jack, queen, king in two suits). Each player receives an initial private card, can bet a fixed amount of 2 chips in the first round, 4 chips in the second round, with a maximum of two raises in each round. A public card is revealed before the second round starts.

In Table \ref{fig:PSROgame} we present such an asymmetric $3 \times 3$ 2-player game generated by the first few epochs of PSRO learning to play Leduc Poker. In the game illustrated here, each player has three strategies that, for ease of the exposition, we call $\{A, B, C\}$ for player 1, and $\{D, E, F\}$ for player 2. Each one of these strategies represents an approximate best response to a distribution over previous opponent strategies. 
In Table \ref{tab:CPtables}  we show the two symmetric counterpart games (see section \ref{sec:theor}) of the empirical game produced by PSRO.

\begin{table}[h!]
\vspace{-0.6cm}
\footnotesize
	\centering
   \begin{game}{3}{3}[][]
   	 & D & E & F  \\
    A & $-2.26,0.02$ & $-2.06,-1.72$ & $-1.65,-1.43$ \\
    B & $-4.77,-0.13$ & $-4.02,-3.54$ & $-5.96,-2.30$  \\
    C & $-2.71,-1.77$ & $-2.52,-2.94$ & $-6.10,1.06$  \\
   \end{game}
\caption{\small Asymmetric PSRO meta game applied to Leduc poker.}
 \label{fig:PSROgame}
 \vspace{-1cm}
\end{table}

\begin{table}[htb]
\vspace{-0.5cm}
\footnotesize
	\centering
\begin{minipage}{.25\textwidth}
   \begin{game}{3}{3}[][]
   	    &  A     &  B   & C  \\
   	 A  &    $-2.26$      & $-2.06$ & $-1.65$\\
   	 B &  $-4.77$ & $-4.02$ & $-5.96$\\
   	 C & $-2.71$ & $-2.52$ & $-6.10$\\
   \end{game}
   \label{fig:CP-PSRO1}
\end{minipage}
\begin{minipage}{.25\textwidth}
   \begin{game}{3}{3}[][]
   	  &  D     &  E   & F  \\
   	 D  &    $0.02$      & $-1.72$ & $-1.43$\\
   	 E &  $-0.13$ & $-3.54$ & $-2.30$\\
   	 F & $-1.77$ & $-2.94$ & $1.06$\\
   \end{game}
   \label{fig:CP-PSRO2}
\end{minipage}
\caption{\small Left - first counterpart game of the PSRO empirical game. Right - second counterpart game of the PSRO empirical game.}\label{tab:CPtables}
\vspace{-0.8cm}
\end{table}

Again we can now analyse the equilbrium landscape of this game, but now using the asymmetric meta-game payoff table and the decomposition result introduced in section \ref{sec:theor}. Since the PSRO meta game is asymmetric we need two populations for the asymmetric replicator equations. Analysing and plotting the evolutionary asymmetric replicator dynamics now quickly becomes very tedious as we deal with two simplices, one for each player. More precisely, if we consider a strategy profile for one player in its corresponding simplex, and that player is adjusting its strategy, this will immediately cause the second simplex to change, and vice versa. Consequently, it is not straightforward anymore to analyse the dynamics. 

\begin{figure}[h!]
 \vspace{-0.6cm}
  \centering
  \begin{minipage}[b]{0.33\textwidth}
     \includegraphics[width=\textwidth]{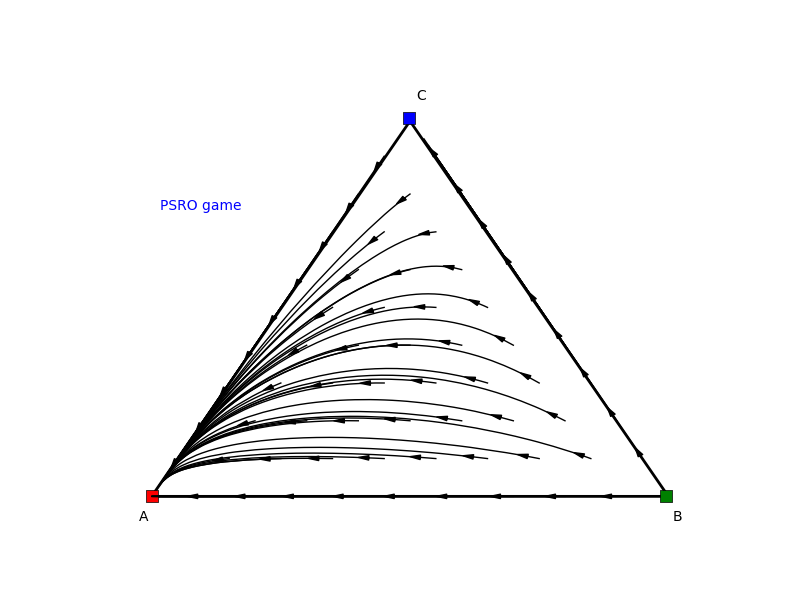}
   \vspace{-1.2cm}
   \caption{\small Trajectory plot of the first CP game.}
    \label{fig:dfieldPSROP1}
  \end{minipage}
  \vspace{-0.4cm}
  \end{figure}

\begin{figure}[h!]
 \vspace{-0.3cm}
  \centering
  \begin{minipage}[b]{0.33\textwidth}
\includegraphics[width=\textwidth]{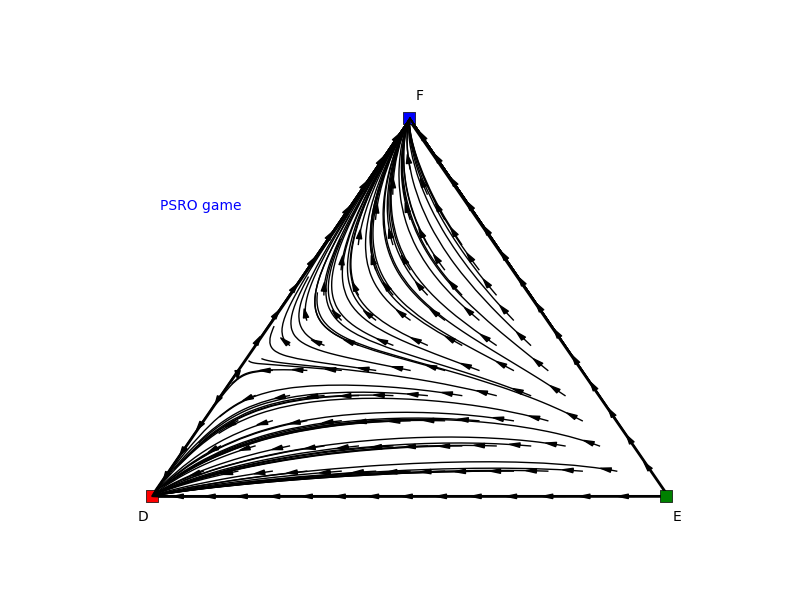}
   \vspace{-1.2cm}
  \caption{\small Trajectory plot of the 2nd CP game.}
    \label{fig:dfieldPSROP2}
  \end{minipage}
 \vspace{-0.4cm}
\end{figure}

In order to facilitate the process of analysing the dynamics we can apply the counterpart theorems to remedy the problem. 
In Figures \ref{fig:dfieldPSROP1} and \ref{fig:dfieldPSROP2} we show the evolutionary dynamics of the counterpart games. As can be observed in Figure \ref{fig:dfieldPSROP1} the first counterpart game has only one equilibrium, i.e., a pure Nash equilibrium in which both players play strategy $A$, which absorbs the entire strategy space. Looking at Figure \ref{fig:dfieldPSROP2} we see the situation is a bit more complex in the second counterpart game, here we observe three equilibiria: one pure at strategy $D$, one pure at strategy $F$, and one unstable mixed equilibrium at the 1-face formed by strategies $D$ and $F$. All these equilibria are Nash in the respective counterpart games\footnote{Banach solver (\url{http://banach.lse.ac.uk/}) is used to check Nash equilibria \cite{Avis10}}. By applying the theory of section \ref{sec:theor} we now know that we only maintain the combination $((1,0,0),(1,0,0))$ as a pure Nash equilibrium of the asymmetric PSRO empirical game, since these strategies have the same support as a Nash equilibrium in the counterpart games. The other equilibria in the second counterpart game can be discarded as candidates for Nash equilibria in the PSRO empirical game since they do not appear as equilibria for player 1. 


Finally, each joint action of the game was estimated with $100$ samples. As the outcome of the game is bounded in the interval $[-13,13]$ we can only guarantee that the Nash equilibrium of the meta game we studied is a $2\epsilon$-Nash equilibrium of the unknown underlying game. It turns out that with $n=100$ and $\epsilon=0.05$, the confidence can only be guaranteed to be above $10^{-8}$. To guarantee a confidence of at least $0.95$ for the same value of $\epsilon=0.05$, we would need at least $n=886 \times 10^3$ samples. 

\vspace{-\secspace cm}
\section{Conclusion}\label{sec:conclusions}
In this paper we have generalised the heuristic payoff table method introduced by Walsh et al. \cite{Walsh02} to two-population asymmetric games. We call such games \textit{meta-games} as they consider complex strategies instead of atomic actions as found in normal-form games. As such they are well suited to investigate real-world multi-agent interactions, as they summarize behaviour in terms of high-level strategies rather than primitive actions. We have shown that a Nash equilibrium of the meta-game is a $2 \epsilon$ Nash equilibrium of the true underlying game, providing theoretical bounds on how much data samples are required to build a reliable meta payoff table. As such our method allows for an equilibrium analysis with a certain confidence that this game is a good approximation of the underlying real game. Finally, we have carried out an empirical illustration of this method in three complex domains, i.e., \textit{AlphaGo}, Colonel Blotto and PSRO, showing the feasibility and strengths of the approach. 

\section*{Acknowledgments}
We wish to thank Angeliki Lazaridou and Guy Lever for insightful comments, the DeepMind AlphaGo team for support with the analysis of the AlphaGo dataset, and Pushmeet Kohli for supporting us with the Colonel Blotto dataset.

\vspace{-\secspace cm}

\bibliographystyle{ACM-Reference-Format}  
\bibliography{sample-bibliography}  

\end{document}